\documentclass[12 pt]{article} 
\usepackage{times,bm}
\usepackage[affil-it,]{authblk}
\usepackage[pagebackref=false,colorlinks=true,linkcolor=blue,citecolor=blue,urlcolor=blue]{hyperref}
\usepackage{geometry}
\usepackage{graphicx}
\usepackage{cite}
\usepackage{amsmath}
\usepackage{amsfonts}
\usepackage{amssymb}
\usepackage{color}
\usepackage{float}
\usepackage{multirow,multicol}
\usepackage{tabulary}
\usepackage{rotating}
\usepackage[flushleft]{threeparttable}
\usepackage{pgfplots,subfigure}
\usepackage{xcolor}
\numberwithin{equation}{section}
\usepackage{tikz}
\usepackage{ulem}
\usepackage{hyperref}
\usepackage{nameref}
\usepackage{comment}

\usepgflibrary{arrows}
\usetikzlibrary{shapes.callouts}
\tikzset{  
	level/.style   = { thick, },
	connect/.style = { dotted, red   },
	notice/.style  = { draw, rectangle callout, callout relative pointer={#1} },
	label/.style   = { text width=2cm }
}

\usepackage{letltxmacro}
\makeatletter
\let\oldr@@t\r@@t
\def\r@@t#1#2{%
	\setbox0=\hbox{$\oldr@@t#1{#2\,}$}\dimen0=\ht0
	\advance\dimen0-0.2\ht0
	\setbox2=\hbox{\vrule height\ht0 depth -\dimen0}%
	{\box0\lower0.4pt\box2}}
\LetLtxMacro{\oldsqrt}{\sqrt}
\renewcommand*{\sqrt}[2][\ ]{\oldsqrt[#1]{#2}}
\makeatother
\begin{document}

\newcommand{{\ri}}{{\rm{i}}}
\newcommand{{\Psibar}}{{\bar{\Psi}}}
\newcommand{\red}{\color{red}}
\newcommand{\blue}{\color{blue}}

\title{
 Lorentz violation in a  family of $(1+2)$-dimensional wormhole
}
\author{\large \textit \textit {Soroush Zare}$^{\ 1}$\footnote{E-mail:soroushzrg@gmail.com}~,~ \textit {Marc de Montigny}$^{\ 2}$\footnote{E-mail:mdemonti@ualberta.ca}~,~ \textit {Hao Chen}$^{\ 3}$\footnote{E-mail:gs.ch19@gzu.edu.cn}~ and~{Hassan Hassanabadi}$^{\ 1}$\footnote{E-mail:h.hasanabadi@shahroodut.ac.ir}  \\

	\small \textit {$^{\ 1}$Faculty of Physics, Shahrood University of Technology, Shahrood, Iran}\\
	\small \textit{P.O. Box 3619995161-316}\\
	\small \textit {$^{\ 2}$ Facult\'e Saint-Jean, University of Alberta, Edmonton, AB T6C 4G9, Canada}\\
	\small \textit {$^{\ 3}$College of Physics, Guizhou University Guiyang 550025, P. R. China}\\
}

\date{}
\maketitle

\begin{abstract}	
 We study neutral Dirac particles confined to a family of $(1+2)$-dimensional wormholes arising from surfaces of revolution with a constant negative Gaussian curvature, in the framework of a comprehensive effective field theory allowing deviations from Lorentz symmetry: the gravitational standard-model extension (SME). The Dirac particles are described with a fixed background tensor field that rules the Lorentz symmetry violation in the CPT-even gauge sector of SME.  We implement this geometrical approach by incorporating non-minimal couplings that possibly induce a Lorentz-symmetry violating term in the modified Dirac equation. We also analyze the exact analytical solutions of the corresponding modified Dirac equation in the presence of a peculiar external magnetic field. 
\end{abstract}

\begin{small}	
Keywords: Standard-model extension (SME); surfaces of revolution; $(1+2)$-dimensional wormhole; Dirac equation.	
\end{small}

\bigskip

\newpage

%%%%%%%%%%%%%%%%%%%%
%%% INTRODUCTION %%%
%%%%%%%%%%%%%%%%%%%%

\section{Introduction}

 In 1916, Flamm, and nearly two decades later, Einstein and Rosen, were the first people that  discussed the idea of the wormhole \cite{Flamm-Einstein-Rosen}.  Wormholes are also known as solutions of Einstein's field equations \cite{Padmanabhan,KunduEPJC2022}. As it is well known, this idea was expanded by many authors. However, as interesting as wormholes may be, the fact remains that none have been observed as of today, so that our motivation for this work stems from its potential applications at the interface between quantum fields in curved space-times and condensed matter physics. For instance, the authors of Ref. \cite{IorioPRD2014} examined various descriptions of deformed graphene offered by different constructions of quantum field theory on curved space-times. Closer to the present paper, let us point out the recent work on $(1+2)$-dimensional wormhole with applications to graphene or graphene wormhole is in Ref. \cite{RojjanasonEPJC1920}.

Also of interest for the present paper is a recent investigation of the spinless stationary Schr\"{o}dinger equation for the electron when it is permanently bound to a generalized Ellis-Bronnikov graphene wormhole-like surface  \cite{deSouza}.   Other works of interest which point toward potential applications of our work include: a complete account of trapped surfaces and the shape of the event horizon for the simplest wormhole topology  \cite{AminneborgCQG1998};  the exact form of the equation of the state of phantom energy that supports the wormhole structure  \cite{JamilIJTP2010};  an investigation of the wormhole with the use of phantom energy as an exotic matter  \cite{DeBenedictis};  wormholes in $(1+2)$-dimensions in different models  \cite{Rahaman};  the partition function of wormhole in $(1+2)$- dimensions \cite{CarlipNPB1990}; the Casimir effects in space-time around the wormhole  \cite{Sorge-Santos-Muniz}, the relationship between the Casimir effect and traversable wormholes \cite{Lima-Sorge}, and the Casimir wormholes in $(1+2)$-dimensions with applications to the graphene \cite{AlencarEPJC2021}.

 Our main objective is to examine neutral Dirac particles confined to various types of wormholes in $(1+2)$ dimensions in a framework allowing deviations from the Lorentz symmetry. As is well known, one may incorporate terms that breaks the Lorentz and CPT symmetries within the standard model of particle physics via a non-trivial theoretical framework the standard model extension (SME) \cite{ColladayPRD9798}.  SME was developed after the observation of spontaneous breaking of Lorentz symmetry in string theories in the first two papers of Ref. \cite{Kostelecky-Samuel-Potting}.  By embedding the Lorentz-symmetry violating terms within all segments of the minimal Standard Model,  the SME provides an approach to investigate physical properties of desired physical systems. The terms leading to the Lorentz-symmetry violation (LSV) are produced as vacuum expectation values of some tensors at high energy scale. In recent years, this approach has been an attractive and convenient platform for much research in different fields of physics \cite{ColladayPRD9798,Kostelecky-Samuel-Potting,Mehlstaubler2022,TassonRPP2014,KosteleckyPRD2004,KosteleckPRLPRD,BetschartNPBPLB2009,CasanaPRD1213}.

 This report of our work comprises the following parts. Section \ref{sec2} sets up the general model of a fermion subject to a wormhole geometry with LSV and CPT-violating terms by means of the SME. We explain how a constant background tensor field governs the LSV in the CPT-even gauge sector of the SME. Our model contains a Maxwell segment, which we reduce to a magnetic field along one direction. In Section \ref{sec3}, we obtain the generalized Dirac equation in a general wormhole space-time, the wormhole being determined for our sakes by a function $R(u)$ which describes the form of the wormhole. In Sections \ref{sec4}, \ref{sec5} and \ref{sec6}, we find the wave-functions and energy eigenvalues for the hyperbolic, elliptic and Beltrami wormholes, respectively, before presenting concluding remarks in Section \ref{sec7}.

%%%%%%%%%%%%%%%%%%%%%%%%%%%%%%%%%%%%%%%%%%%%%%%%

%%% SEC 2

%%%%%%%%%%%%%%%%%%%%%%%%%%%%%%%%%%%%%%%%%%%%%%%%

\section{ Lorentz-violation model and wormhole geometry
 \label{sec2}}
 
The conventional standard model can be extended so as to incorporate gravitation as well as Lorentz-symmetry violating (LSV) and CPT-violating terms; this more extensive model is called the standard-model extension (SME) \cite{ColladayPRD9798,TassonRPP2014}. Likewise,  we will  analyze some aspects of the fermion sector of quantum field theory in the context of SME.  The fermionic part of the action in flat space-time, which contains the QED Lagrangian modified under the influence of an even sector of SME, is
\begin{equation}\label{Action}
\begin{split}
&\mathcal{S}_{\Psi}=\int \mathrm{d}^{4}x \mathcal{L}_{\mathrm{modQED}}.
\end{split}
\end{equation}
 We can propose the corresponding $(1+3)$-dimensional QED Lagrangian density as a sum of modified Dirac and Maxwell Lagrangian densities to describe the quantum dynamics of the non-gravitational fermion field, $\Psi$, under the influence of the CPT-even and LSV non-minimal coupling, \cite{KosteleckyPRD2004}
\begin{equation}\label{LagrDensity1}
\begin{split}
\mathcal{L}_{\mathrm{modQED}} & =\mathcal{L}_{\mathrm{modDirac}}+\mathcal{L}_{\mathrm{Maxwell}}\\
& = \frac{\mathrm{i}}{2}\left(\bar{\Psi}\gamma^{a}\nabla_{a}\Psi-\left(\bar{\Psi}\bar{\nabla}_{a}\right)\gamma^{a}\Psi\right)-M\bar{\Psi}\Psi-\frac{1}{4}F_{ab}F_{cd}\eta^{ac}\eta^{bd}. 
\end{split}
\end{equation}
The modified Dirac Lagrangian  describes  a neutral fermion of mass $M$ subject to a background tensor field  at a high-energy scale that rules the Lorentz symmetry violation in the CPT-even electrodynamics of the SME. This Lagrangian is written in terms of the covariant derivative with relevant non-minimal coupling as \cite{KosteleckPRLPRD,BetschartNPBPLB2009,CasanaPRD1213}
\begin{equation}\label{CovDeriv}
	\nabla_{a}=\partial_{a}+\frac{\lambda}{2}\left(K_{F}\right)_{abcd}\gamma^{b}F^{cd}.
\end{equation}
Throughout this work, we take natural units with $c = \hbar = G_{N} = 1$.  Latin indices stand for local Lorentz indices and run from 0 to 3, whereas Latin indices with tilde mark run from 1 to 3. The metric tensor in Minkowski space-time is indicated by $\eta_{ab}$ with
signature $(+---)$.  $\lambda$ denotes the coupling constant, and $\gamma^{b}$ are the standard Dirac matrices in Minkowski space-time. $F_{a b}\equiv\partial_{a}A_{b}-\partial_{b}A_{a}$ is the the electromagnetic field tensor, such that $F^{cd}=\eta^{a c}\eta^{d b}F_{a b}$.  

The factor $(K_F)_{abcd}$ denotes the constant background tensor field that governs the LSV in the CPT-even gauge sector of SME.  The conservation of energy and momentum corresponds to constant $(K_F)_{abcd}$   (see the 2004 paper in Ref. \cite{KosteleckPRLPRD} and Ref. \cite{CasanaPRD1213}.) 
This tensor  consists of nineteen independent components with  special properties \cite{KosteleckPRLPRD,BetschartNPBPLB2009,CasanaPRD1213}. It has symmetries similar to the Riemann curvature tensor with a double-trace condition (zero double-trace), that is
\begin{equation}\label{tersor2}
(K_F)_{abcd} = (K_F)_{[ab][cd]}, \quad (K_F)_{abcd} = (K_F)_{cdab}, \quad (K_F)^{ab}_{\,\,\,\,\,\,ab} = 0.
\end{equation}
 In order to introduce LSV, we adopt appropriate linear combinations between $(K_F)_{abcd}$ and four rank-2 tensors $k_{DE}$, $k_{HB}$, $k_{DB}$ and $k_{HE}$ as follows \cite{KosteleckPRLPRD}
\begin{equation}\label{tersor3}
\begin{split}
&(k_{DE})_{jk} = -2 (K_{F})_{0j0k}, \\
&(k_{HB})_{jk} = \frac{1}{2} \epsilon_{jpq}\,\epsilon_{klm} (K_{F})_{pqlm}, \\
&(k_{DB})_{jk} = -(k_{HE})_{kj} = \epsilon_{kpq}(K_{F})_{0jpq}. 
\end{split}
\end{equation} 
(The subscripts $E$, $B$, $D$, $H$ refer to the electric, magnetic, displacement and induction fields and exploit an analogy between LSV electrodynamics in vacuum and homogeneous anisotropic media \cite{KosteleckPRLPRD}.) The $(K_F)_{abcd}$ tensor  controls the LSV and, as could be expected, its values is small.
In addition, the $(K_F)_{abcd}$ tensor is expressed as two parity sectors, such that the tensors $k_{DE}$ and $k_{HB}$ possess even parity, while $k_{DE}$ and $k_{HB}$ have odd parity.

 Eqs. \eqref{LagrDensity1} and \eqref{CovDeriv} show that the Lagrangian of the modified Maxwell theory contains the standard Maxwell  and an additional LSV and CPT-even term in Minkowski space-time that can be written as
\begin{equation}\label{ModMaxLag}
\mathcal{L}_{modMaxwell} = -\frac{1}{4}F_{ab}F_{cd}\eta^{ac}\eta^{bd}  +\frac{\lambda}{2}\left(K_{F}\right)_{abcd}\bar{\Psi}\sigma^{ab}\Psi F^{cd}.
\end{equation}
Thus, the Dirac equation, modified by the covariant derivative $\nabla_{a}$  with a CPT-even non-minimal coupling term, can be written in Minkowski space-time as
\begin{equation}\label{LSVDiracEqFlat}
\left[\mathrm{i}\gamma^{a}\partial_{a}+\frac{\lambda}{2}(K_F)_{abcd} \sigma^{ab} F^{cd}-M\right]\Psi\left(t,{\bf r}\right)=0,
\end{equation}
in which the $\sigma^{ab}$ operator is given by $\sigma^{ab}=\mathrm{i}[\gamma^{a},\gamma^{b}]/2$, whose components read $\sigma^{0\tilde{i}}=\mathrm{i}\alpha^{\tilde{i}}$ and $\sigma^{\tilde{i}\tilde{j}}=\epsilon_{\tilde{i}\tilde{j}\tilde{k}}\Sigma^{\tilde{k}}$, where $\alpha^{\tilde{i}}$ and $\Sigma^{\tilde{k}}$ are matrices defined in terms of Pauli matrices that can be used to rewrite the Dirac matrices as
\begin{equation}
\gamma^{0}=\beta=
\begin{pmatrix}
1 & 0 \\
0 & -1
\end{pmatrix}, \quad
\gamma^{\mathrm{i}}=
\begin{pmatrix}
0 & \sigma^{\mathrm{i}} \\
-\sigma^{\mathrm{i}} & 0
\end{pmatrix}, \quad
\alpha^{\mathrm{i}}=
\begin{pmatrix}
0 & \sigma^{\mathrm{i}} \\
\sigma^{\mathrm{i}} & 0
\end{pmatrix}, \quad
\Sigma^{\tilde{k}}=
\begin{pmatrix}
\sigma^{\tilde{k}} & 0 \\
0 & \sigma^{\tilde{k}}
\end{pmatrix}.
\end{equation}
The extended form of the second term of Eq. \eqref{LSVDiracEqFlat}, using Eq. \eqref{tersor3} and  keeping in mind  that $F^{0\tilde{i}}=-F^{\tilde{i}0}=-E^{\tilde{i}}$ and $F^{\tilde{i}\tilde{j}}=-\epsilon^{\tilde{i}\tilde{j}\tilde{k}}B_{\tilde{k}}$, can be written in terms of the electric and magnetic fields as \cite{CasanaPRD1213}
\begin{equation}\label{SecTermDirac}
\begin{split}
\frac{\lambda}{2}(K_F)_{abcd} \sigma^{ab} F^{cd} &= \lambda \sigma^{0\tilde{i}} \left[\left(k_{DE}\right)_{\tilde{i}\tilde{j}}E^{\tilde{j}}+\left(k_{DB}\right)_{\tilde{i}\tilde{j}}B^{\tilde{j}}\right]\\&+
\frac{\lambda}{2}\epsilon_{\tilde{i}\tilde{j}\tilde{k}} \sigma^{\tilde{i}\tilde{j}} \left[\left(k_{HE}\right)_{\tilde{k}\tilde{q}}E^{\tilde{q}}+\left(k_{HB}\right)_{\tilde{k}\tilde{q}}B^{\tilde{q}}\right].
\end{split}
\end{equation}
From now on,  we consider only one non-zero component of the tensor $(K_F)_{abcd}$,  which we take to be  $(k_{DB})_{21}$. Thus, Eq. \eqref{SecTermDirac} can be reduced to 
\begin{equation}\label{SecTermDirac2}
\frac{\lambda}{2}(K_F)_{abcd} \sigma^{ab} F^{cd} = \lambda \sigma^{02} \left(k_{DB}\right)_{21}B^{1},
\end{equation}
in which $B^{1}$ is an external magnetic field configuration in the $x$-direction. 

Eq. \eqref{SecTermDirac2}  shows that keeping $\left(k_{DB}\right)_{21}$ as only one non-zero constant component of the tensor field $(K_F)_{abcd}$ leads to having an external magnetic field configuration $B^{1}$ in $(1+3)$-dimensional flat space-time so that multiplying $\left(k_{DB}\right)_{21}$ by $B^{1}$ gives rise to a particular  LSV and CPT-even non-minimal coupling related to a spinor field.   In the following,  our motivation is to investigate a spinor approach in general relativity subject to the covariant derivative containing that special case of the  LSV and CPT-even non-minimal coupling in the presence of curved space-time generated by a $(1+2)$-dimensional wormhole.

 Now let us turn to a type of surface of revolution with constant Gaussian curvature  that can be seen as a surface produced by a profile-curve placed in the plane $(x,z)$ that rotates around the $z$-axis by an angle as much as $2\pi$ \cite{Eisenhart-Spivak,IorioPRD2014}. Hence, this surface of revolution evokes a wormhole-type space-time with two event horizons. Geometrically, the coordinate of a point on such a surface can be represented by the following position vector,   
\begin{equation}\label{Pvector}
{\bf r}(u,v) = x(u,v)\hat{i}+y(u,v)\hat{j}+z\hat{k},
\end{equation}
placed on the surface parametrized in 3-dimensional space, with
\begin{equation}\label{surfacepara}
x(u,v) = R(u) \mathrm{cos}v, \quad y(u,v) = R(u) \mathrm{sin}v,\quad z(u)=\pm\int^{u}\sqrt{1-R'(u)^{2}}\,\,\mathrm{d}u,
\end{equation}
where the $u$ and $v$ parameters are the new spatial coordinates.  These coordinates $u$ and $v$ play the role of  meridian and angular coordinates, respectively, of a $(1+2)$-dimensional wormhole cropping up when the structure of space-time is curved by gravity. The form of the wormhole can be characterized by the function $R(u)$ \cite{Eisenhart-Spivak,IorioPRD2014,RojjanasonEPJC1920}. The relation between $z(u)$ and $R(u)$ is  given by the condition  $z'^2(u)+R'^2(u)=1$. Therefore,  by implementing Eq. \eqref{surfacepara} into the line element of $(1 + 3)$-dimensional flat space-time, we describe a reduced space-time with the following line element corresponding to the $(1+2)$-dimensional wormhole:
\begin{equation}\label{wormmetric}
ds^2=dt^2-du^2-R^{2}(u)dv^2.
\end{equation} 
 Such a dimensional reduction of space-time may leads to the appearance of gravitational effects due to the wormhole  \cite{IorioPRD2014,RojjanasonEPJC1920}. 
It should be noted that the constant Gaussian curvature  is given by \cite{Eisenhart-Spivak,IorioPRD2014}
\begin{equation}\label{constGausscurvature}
\mathcal{K} = -\frac{R''(u)}{R(u)}.
\end{equation}
Obviously, the above equation becomes similar to the harmonic oscillator differential's equation, whose solutions can depend on the sign of $\mathcal{K}$.

 We are interested in models that describe the LSV and the difference between local Lorentz transformations and general coordinate transformations without any torsion in the background, relying on the tetrad formalism adjusted for such theoretical models \cite{KosteleckPRLPRD,BetschartNPBPLB2009}.
Accordingly, the corresponding constant background tensor field can be adjusted appropriately to curved space-time as
\begin{equation}\label{LVtensorCurved}
	\left(K_{F}\right)_{\mu\nu\alpha\beta} = e^{a}_{\,\,\,\mu} e^{b}_{\,\,\,\nu} e^{c}_{\,\,\,\alpha} e^{d}_{\,\,\,\beta} \left(K_{F}\right)_{abcd}.
\end{equation}
\begin{comment}
In addition, Eq. \eqref{tersor3} must be well adapted to curved space-time
\begin{equation}\label{tersor3curved}
\begin{split}
&(k_{DE})_{\nu\beta} = e^{j}_{\,\,\,\nu} e^{k}_{\,\,\,\beta} (k_{DE})_{jk} = -2 e^{0}_{\,\,\,\mu} e^{j}_{\,\,\,\nu} e^{0}_{\,\,\,\alpha} e^{k}_{\,\,\,\beta} (K_{F})_{0j0k}, \\
&(k_{HB})_{\nu\beta} = e^{j}_{\,\,\,\nu} e^{k}_{\,\,\,\beta} (k_{HB})_{jk} = \frac{1}{2} \epsilon_{jpq}\,\epsilon_{klm}\,\,e^{p}_{\,\,\,\mu} e^{q}_{\,\,\,\nu} e^{l}_{\,\,\,\alpha} e^{m}_{\,\,\,\beta} (K_{F})_{pqlm}, \\
&(k_{DB})_{\nu\beta} = e^{j}_{\,\,\,\nu} e^{k}_{\,\,\,\beta} (k_{DB})_{jk} = \epsilon_{kpq}\,\,e^{0}_{\,\,\,\mu} e^{j}_{\,\,\,\nu} e^{p}_{\,\,\,\alpha} e^{q}_{\,\,\,\beta}(K_{F})_{0jpq},\\
&(k_{HE})_{\nu\beta} = e^{j}_{\,\,\,\nu} e^{k}_{\,\,\,\beta}(k_{HE})_{kj} = -\epsilon_{kpq}\,\,e^{0}_{\,\,\,\mu} e^{j}_{\,\,\,\nu} e^{p}_{\,\,\,\alpha} e^{q}_{\,\,\,\beta}(K_{F})_{0jpq}. 
\end{split}
\end{equation}
where Eq. \eqref{tersor3curved} is written in terms of the only non-zero component of the constant background tensor field, that is, $(k_{DB})_{21}$.
\end{comment}

%%%%%%%%%%%%%%%%%%%%%%%%%%%%%%%%%%%%%%%%%%%%%%%%

%%% SEC 3

%%%%%%%%%%%%%%%%%%%%%%%%%%%%%%%%%%%%%%%%%%%%%%%%

\section{Fermions in a $(1+2)$-dimensional wormhole with  LSV and CPT-even non-minimal coupling \label{sec3}}

As mentioned, we study relativistic fermions in the curved space-time of a wormhole's surface with the background tensor field that rules the LSV in the CPT-even gauge sector of the SME.  In order to achieve this, we add the Einstein-Hilbert action $\mathcal{S}_{EH}$ with the action of Eq. \eqref{Action} on a curved space-time so that we utilize the action \cite{KosteleckyPRD2011}
\begin{equation}\label{ActionWorm}
\begin{split}
&\mathcal{S}_{EH}+\mathcal{S}_{\Psi,curved}=\int \mathrm{d}^{3}x\,\,\sqrt{-g} \left(\mathcal{L}_{\mathrm{EH}}+\mathcal{L}_{\mathrm{modQED,curved}}\right),\\
&\mathcal{L}_{\mathrm{EH}}=\frac{1}{16\pi}\left(\mathcal{R}-2\Lambda\right),\\
&\mathcal{L}_{\mathrm{modQED,curved}} = \frac{\mathrm{i}}{2}\left(\bar{\Psi}\gamma^{\mu}\mathcal{D}_{\mu}\Psi-\left(\bar{\Psi}\bar{\mathcal{D}}_{\mu}\right)\gamma^{\mu}\Psi\right)-M\bar{\Psi}\Psi-\frac{1}{4}F_{\mu\nu}F_{\alpha\beta}g^{\mu\alpha}g^{\nu\beta}. 
\end{split}
\end{equation}
 The symbol $g$ is the determinant of the metric tensor $g_{\mu\nu}$ of the line element \eqref{wormmetric}, $G$ is Newton's gravitational constant,  and $\Lambda$  the cosmological constant. 
The symbol $\mathcal{R}$ stands for the Ricci curvature scalar, the value of which depends on $g_{\mu\nu}$ and,  here, $\mathcal{R}=-2R''(u)/R(u)$. Besides, the generalized Dirac matrices are indicated by $\gamma^{\mu}$ given by $\gamma^{\mu} = e^{\mu}_{\,\,\,a}\gamma^{a}$, where $e^{\mu}_{\,\,\,a}$  are the inverse tetrads $e^{a}_{\,\,\,\mu}$, and also with regard to the signature in Eq. \eqref{wormmetric} the standard Dirac matrices $\gamma^{a}$ can be determined in terms of Pauli matrices as $\gamma^{0} = \sigma^{3}$, $\gamma^{1} = \mathrm{i}\sigma^{1}$ and $\gamma^{2} = \mathrm{i}\sigma^{2}$. By the way, this choice of $\gamma^{a}$ satisfy the anticommutation relation $\{\gamma^{a},\gamma^{b}\} = 2\eta^{ab}$, with $\eta^{ab} = (+,-,-)$. Moreover, to find the components of $\gamma^{\mu}$, we need to have a appropriate choice of tetrads $e^{a}_{\,\,\,\mu}$ whose inverse becomes $e^{t}_{\,\,\,0} = e^{u}_{\,\,\,1} = 1$ and $e^{v}_{\,\,\,2} = 1/R(u)$. Hence, the components of $\gamma^{\mu}$ can be written as $\gamma^{t} = \gamma^{0}$, $\gamma^{u} = \gamma^{1}$ and $\gamma^{v} = \gamma^{2}/R(u)$. 

The covariant derivative $\mathcal{D}_{\mu}$ appearing in Eq. \eqref{ActionWorm} contains a possible scenario of the LSV and CPT-even non-minimal coupling, defined in the previous section, as well as the gravitationally coupled SME. Therefore, with the space-time reduction from $(1+3)$ to $(1+2)$ dimensions, we can express $\mathcal{D}_{\mu}$ as follows:
\begin{equation}\label{CovDerivWorm}
\mathrm{i}\gamma^{\mu}\mathcal{D}_{\mu}=\mathrm{i}\gamma^{\mu}\partial_{\mu}+\mathrm{i}\gamma^{\mu}\Gamma_{\mu}+\lambda \sigma^{tv} \left(k_{DB}\right)_{vu}B^{u},  
\end{equation}
where
\begin{equation}
\begin{split}
&\Gamma_{\mu}=-\frac{\mathrm{i}}{4}\omega_{\mu ab}\sigma^{ab}, \quad	
\sigma^{tv} = \frac{\mathrm{i}}{2}[\gamma^{t},\gamma^{v}]\equiv e^{t}_{\,\,\,0} e^{v}_{\,\,\,2}\sigma^{02},\\
&(k_{DB})_{vu} = e^{2}_{\,\,\,v} e^{1}_{\,\,\,u} (k_{DB})_{21}, \quad B^{u}\rightarrow B(u)\hat{u}
\end{split}
\end{equation}
Eq. \eqref{CovDerivWorm} is written in terms of the only one non-zero component of the constant background tensor field, that is, $(k_{DB})_{21}$. 
The affine connection is $\Gamma_{\mu}$, in which the spin connection is denoted by $\omega_{\mu ab}$ and can be obtained from the Maurer-Cartan structure equations, $\mathrm{d}\hat{\theta}^{a}+\omega^{a}_{\,\,\,b}\wedge\hat{\theta}^{b} = 0$, which are written in the absence of torsion, with $\omega^{a}_{\,\,\,b} = \omega^{\,\,a}_{\mu\,\,\,b}\mathrm{d}x^{\mu}$. Thus, the non-null components of the spin connection $\omega_{\mu ab} = \eta_{ac}\, \omega^{\,\,c}_{\mu\,\,\,b}$ and the affine connection $\Gamma_{\mu}$, for the metric \eqref{wormmetric}, are  $\omega_{v 21} = -\omega_{v 12} = -R'(u)$ and $\Gamma_{v} = -\mathrm{i}\sigma^{3} R'(u)/2$, respectively. Thereby, the $\mathrm{i}\gamma^{v}\Gamma_{v}$ combination gets $-\sigma^{1}R'(u)/2R(u)$.
 Moreover, the external magnetic field $B^{u}$ is written in the general reference frame. 

Our modified Dirac equation, which  stems  from embedding the non-minimal coupling \eqref{CovDerivWorm}, can be written as 
\begin{equation}\label{ModDirac1}
\left[\gamma^{0}\partial_{t}+\gamma^{1}\left(\mathrm{i}\partial_{u}+\mathrm{i}\frac{R'(u)}{2R(u)}+\lambda\left(k_{DB}\right)_{21}B(u) \right)+\mathrm{i}\frac{\gamma^{2}}{R(u)}\partial_{v}-M\right]\Psi(t,{\bf{r}})=0.
\end{equation}
 Let us write the solution Eq. \eqref{ModDirac1}  as
\begin{equation}\label{PrimWF}
\Psi(t,{\bf{r}}) = e^{-\mathrm{i}Et+\mathrm{i}mv}
\begin{pmatrix}
\psi_{1}(u)\\
\psi_{2}(u)
\end{pmatrix},
\end{equation}
in which $E$ denotes the relativistic energy of the system,  $\psi_{1}(u)$ and $\psi_{2}(u)$ are called two-component spinors, and  $m=\ell+\frac12$, with $\ell = 0, \pm 1,\pm 2,\dots$ are the eigenvalues of the total angular momentum operator $\hat{J}_{z} = -\mathrm{i}\partial_{v}$.
 Thus, Eq. \eqref{ModDirac1} can be written as coupled equations in terms of $\psi_{1}(u)$ and $\psi_{2}(u)$: 
\begin{subequations}\label{CoupledDiracEq}
\begin{align}
& \left[E-M\right]\psi_{1}(u)+\left[-\partial_{u}-\frac{R'(u)}{2R(u)}+\mathrm{i}\lambda\left(k_{DB}\right)_{21}B(u)-\frac{m}{R(u)}\right]\psi_{2}(u) = 0,\label{CoupledDiracEq1}\\
&\left[E+M\right]\psi_{2}(u)+\left[\partial_{u}+\frac{R'(u)}{2R(u)}-\mathrm{i}\lambda\left(k_{DB}\right)_{21}B(u)-\frac{m}{R(u)}\right]\psi_{1}(u)= 0. \label{CoupledDiracEq2}
\end{align}
\end{subequations}
We obtain second-order differential equation in terms of $\psi_{1}(u)$ by eliminating $\psi_{2}(u)$ from Eq. \eqref{CoupledDiracEq2} and replacing into Eq. \eqref{CoupledDiracEq1}:
\begin{equation}\label{CoupledDiracEq3}
\begin{split}	
&\psi_{1}''(u)-\left[2\mathrm{i}\lambda\left(k_{DB}\right)_{21}B(u)-\frac{R'(u)}{R(u)}\right]\psi_{1}'(u)+\left[k^2-\frac{m^2}{R(u)^2}+\frac{mR'(u)}{R(u)^2}-\frac14\frac{R'(u)^2}{R(u)^2}\right.\\
&\left.-\left(\lambda\left(k_{DB}\right)_{21}\right)^2B(u)^2 -\mathrm{i}\lambda\left(k_{DB}\right)_{21}\frac{B(u)R'(u)}{R(u)}-\mathrm{i}\lambda\left(k_{DB}\right)_{21}\partial_{u}B(u)+\frac{R''(u)}{2R(u)}
\right]\psi_{1}(u) = 0,
\end{split}
\end{equation}
where $k^2 = E^2 - M^2$. Hereafter, in order to find analytical, exact solutions of  Eq. \eqref{CoupledDiracEq3}, we specify the external magnetic fields $B(u)$ as well as the form of the wormhole via the meridian function $R(u)$ associated with a particular solution of Eq. \eqref{constGausscurvature}.

 We consider the magnetic field defined in terms of $R(u)$ of Eq. \eqref{surfacepara} so that $B(u)= B_{0}R'(u)/R(u)$, where $B_{0}$ is a constant. 
Hence, Eq. \eqref{CoupledDiracEq3} can be rewritten as 
\begin{equation}\label{DiracEq4}
\begin{split}	
&\psi_{1}''(u)+\left[2\tau+1\right]\frac{R'(u)}{R(u)}\psi_{1}'(u)+\left[k^2-\frac{m^2}{R(u)^2}+\frac{mR'(u)}{R(u)^2}+\left(\tau^2-\frac14\right)\frac{R'(u)^2}{R(u)^2}\right.\\
&\left.+\left(\tau+\frac12\right)\frac{R''(u)}{R(u)}
\right]\psi_{1}(u) = 0,  \qquad \qquad\qquad \tau=-\mathrm{i}\lambda\left(k_{DB}\right)_{21}B_{0}.
\end{split}
\end{equation}
In order to further our study, we examine various types of $R(u)$-functions and consider the sign of the constant Gaussian curvature $\mathcal{K}$  (given in Eq. \eqref{constGausscurvature}) that can determine the nature of the surfaces of revolution.
To begin with, let us return to Eq. \eqref{constGausscurvature} and  consider the form 
\begin{subequations}
\begin{align}
&R(u) =
b_{1}\,\mathrm{sinh}\left(\frac{u}{r}\right)+b_{2}\,\mathrm{cosh}\left(\frac{u}{r}\right), \quad \text{for} \quad \mathcal{K} = -\frac{1}{r^2}, \label{minusK} \\
&R(u) = d\, \mathrm{cos} \left(\frac{u}{r}+\phi\right), \qquad \qquad \qquad \,\, \text{for} \quad \mathcal{K} = \frac{1}{r^2}, \label{plusK}
\end{align}
\end{subequations}  
where $b_{1}$, $b_{2}$, $d$ and $r$ are constant, and $\phi$ plays the role of a phase of the oscillation. Accordingly, all the surfaces of revolution stemming from Eq. \eqref{minusK} 
can be classified as the hyperbolic pseudosphere surface (hyperbolic wormhole), elliptic pseudosphere surface (elliptic wormhole), and Beltrami pseudosphere surface (Beltrami wormhole). 

This implies that (1) with $b_{1} = 0$,  the meridian function is $R(u) = b_{2}\, \mathrm{cosh}\left(\frac{u}{r}\right)$, describing the hyperbolic wormhole,  (2)  if  $b_{2} = 0$, we  have $R(u) = b_{1}\, \mathrm{sinh}\left(\frac{u}{r}\right)$, for the elliptic wormhole, and (3) with $b_{1}=b_{2} \equiv b$, we find $R(u) = b\, e^{\frac{u}{r}}$, with a Beltrami wormhole \cite{Eisenhart-Spivak,IorioPRD2014}. Hereupon, we plot the wormholes above for more details in the following.
\begin{figure}[H]
	\centering
\subfigure[With $b_{2}=1$ and $r=1$.]{%
\label{fig11}%
\includegraphics[height=6cm,width=7cm]{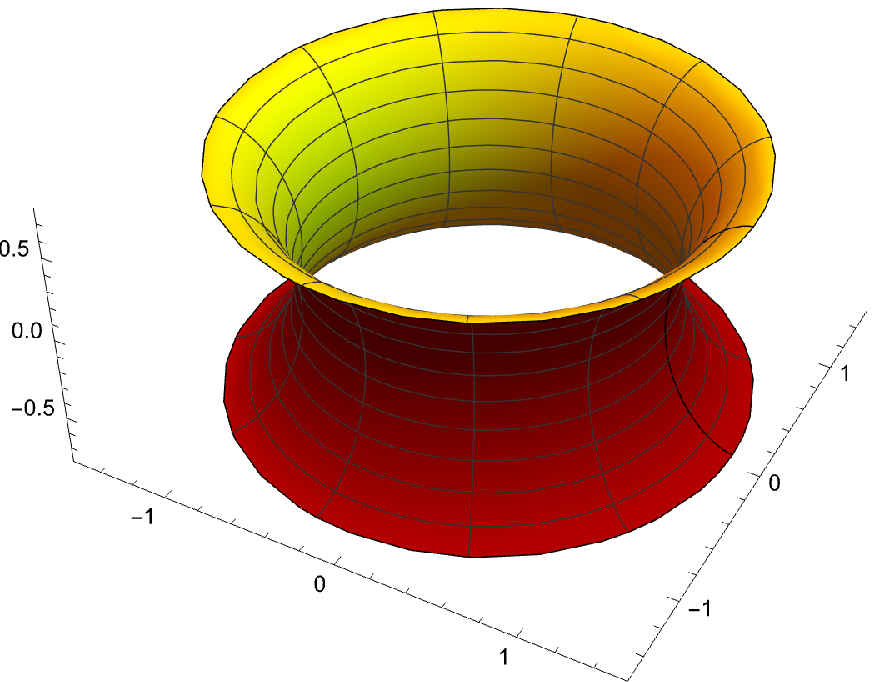}}%
\qquad
\subfigure[With $b_{2}=0.1$ and $r=1$.]{%
\label{fig12}%
\includegraphics[height=6cm]{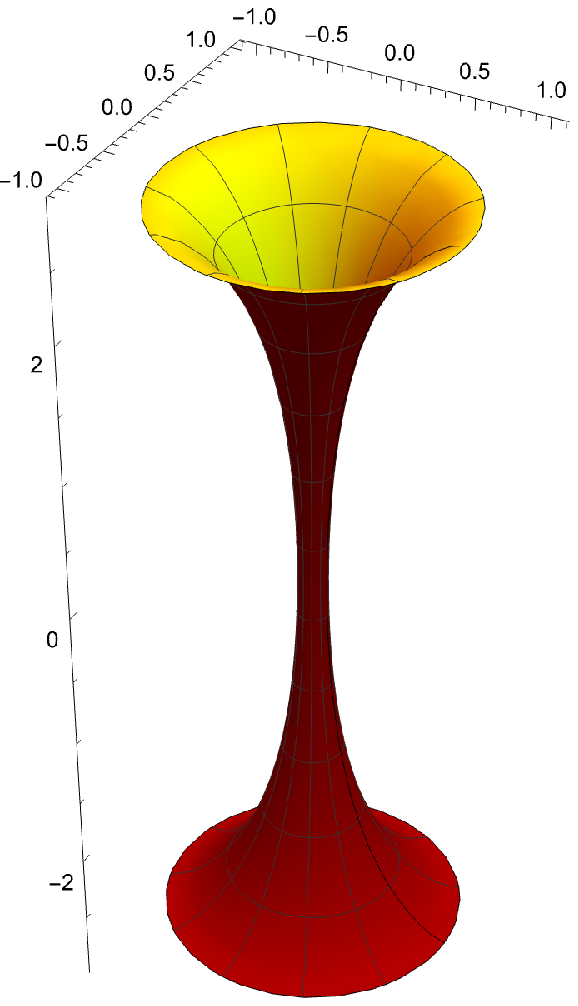}}%
\caption{The hyperbolic wormhole is plotted by taking $R(u)= b_{2}\, \mathrm{cosh}\left(\frac{u}{r}\right)$ (whose range is $R(u) \in \left[b_{2},\sqrt{b_{2}^{2}+r^2}\right]$)  
so that
	$x(u,v) = b_{2}\, \mathrm{cosh}\left(\frac{u}{r}\right) \mathrm{cos} (v)$, $y(u,v) = b_{2}\, \mathrm{cosh}\left(\frac{u}{r}\right) \mathrm{sin}(v)$ and $z(u) =-\mathrm{i} r\, E(\frac{\mathrm{i} u}{r},\frac{b_{2}^2}{r^2})$ (which is the symbol of the elliptic integral of the second kind for $u \in \left[-\mathrm{cosh}^{-1}\left(\sqrt{1+\frac{r^2}{b_{2}^{2}}}\right),\mathrm{cosh}^{-1}\left(\sqrt{1+\frac{r^2}{b_{2}^{2}}}\right)\right]$) with $v\in [0,2\pi]$.	
}\label{fig1}
\end{figure}
In selecting the values of $u$ to plot Fig. \ref{fig1}, the $z\in \mathbb{R}$ condition must be satisfied. By the way, as the $\frac{b_{2}}{r}$ value decreases, the hyperbolic wormhole stretches along its axis (the $z$-axis direction).

\begin{figure}[H]
\centering
\subfigure[With $\varphi=\frac{\pi}{4}$ and $r=1$.]{%
\label{fig21}%
\includegraphics[height=6cm,width=7cm]{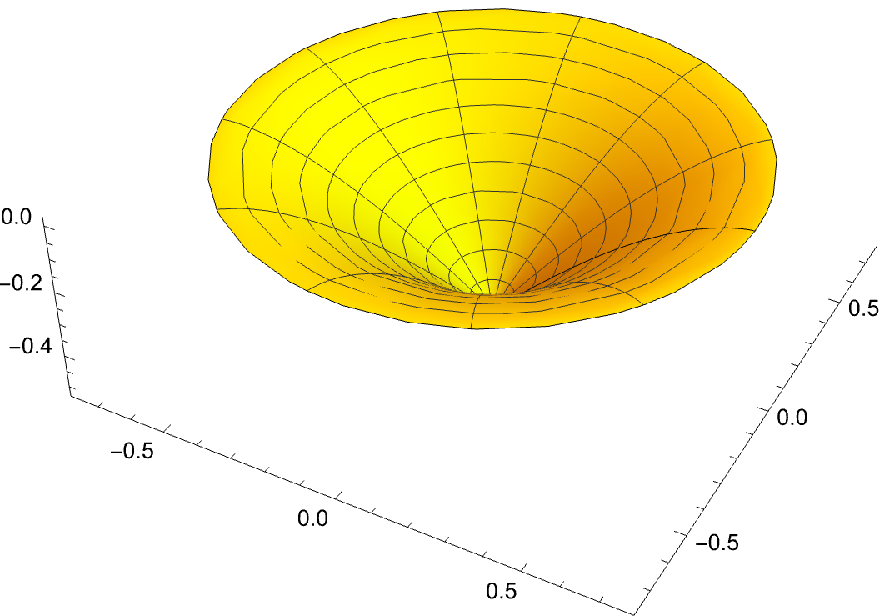}}%
\qquad
\subfigure[With $\varphi=\frac{\pi}{16}$ and $r=1$.]{%
\label{fig22}%
\includegraphics[height=6cm]{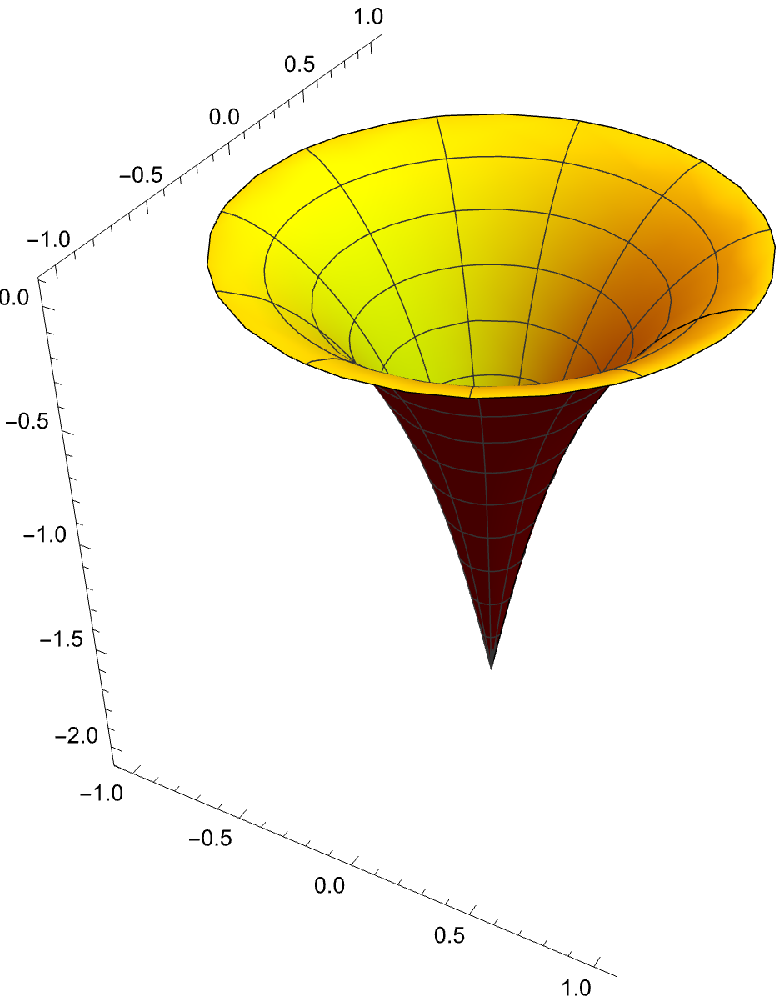}}%
\caption{
The elliptic wormhole is plotted by taking $R(u)= b_{1}\, \mathrm{sinh}\left(\frac{u}{r}\right)$ (whose range is $R(u) \in \left[0,b_{1}\right]$ with $b_{1}=r\mathrm{cos}\varphi<r$)  
so that
$x(u,v) = b_{1}\, \mathrm{sinh}\left(\frac{u}{r}\right) \mathrm{cos} (v)$, $y(u,v) = b_{1}\, \mathrm{sinh}\left(\frac{u}{r}\right) \mathrm{sin}(v)$ and $z(u) =\pm\int\sqrt{1-\left(R'(u)\right)^2} du$. The range of $u$ and $v$ are given by 
$u \in \left[0,\mathrm{sinh}^{-1}\left(\mathrm{cot}\varphi\right)\right]$) with $v\in [0,2\pi]$.		
}\label{fig2}
\end{figure}
\begin{figure}[H]
\centering
\subfigure[With $b=1$, $r=1$ and $-3.37\le u  \le 0$]{%
\label{fig31}%
\includegraphics[height=6cm,width=7cm]{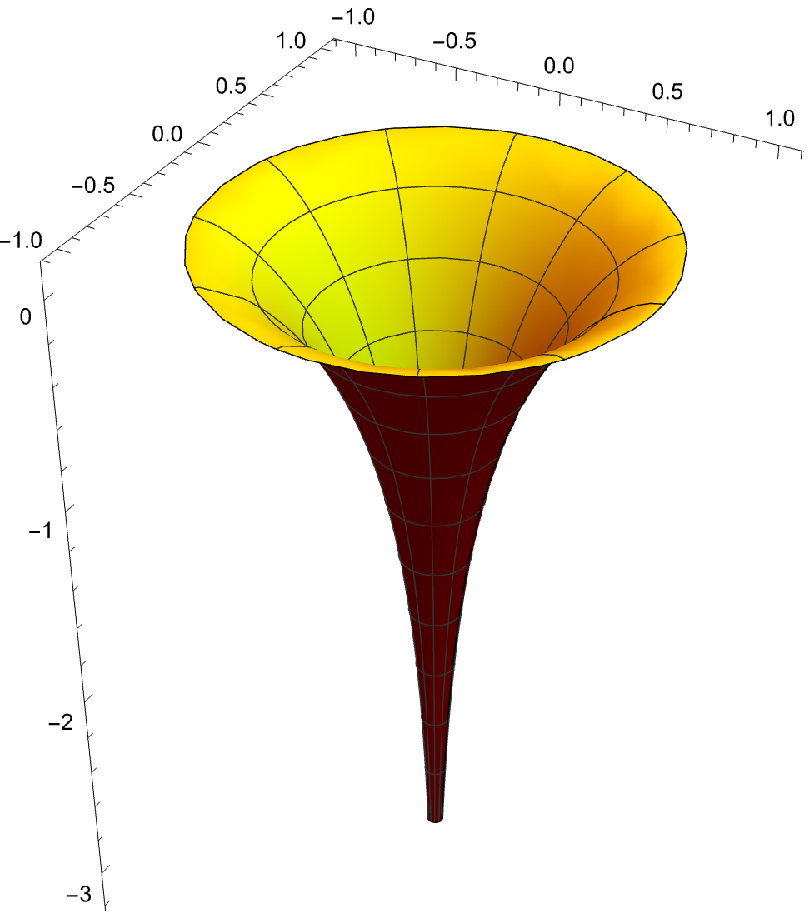}}%
\qquad
\subfigure[With $b=2.7$, $r=1$ and $-3.37\le u  \le \mathrm{ln}(\frac{1}{2.7})$]{%
\label{fig32}%
\includegraphics[height=6cm]{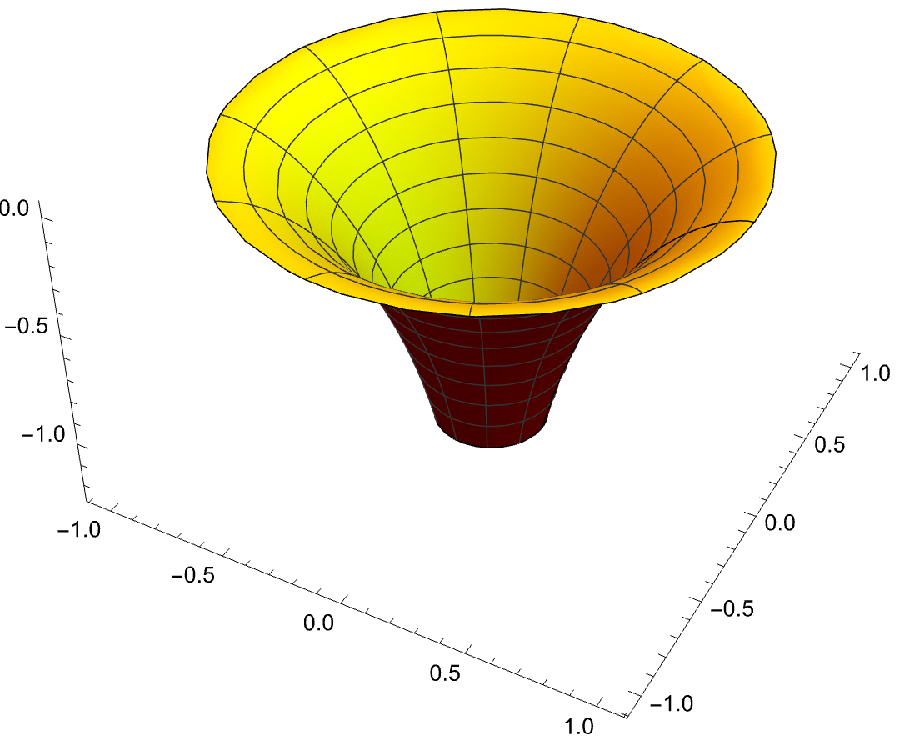}}%
\caption{
The Beltrami wormhole is plotted by taking $R(u)= b\, e^{\frac{u}{r}}$ (whose range is $R(u) \in \left[0,r\right]$)  
so that
$x(u,v) = b\, e^{\frac{u}{r}} \mathrm{cos} (v)$, $y(u,v) = b\, e^{\frac{u}{r}} \mathrm{sin}(v)$ and $z(u) =\pm\int\sqrt{1-\left(R'(u)\right)^2} du$. The range of $u$ and $v$ are given by 
$u \in \left[-\infty,r \mathrm{ln}(\frac{r}{b})\right]$) with $v\in [0,2\pi]$.			
}\label{fig3}
\end{figure}

Corresponding to Fig. \ref{fig2}, the range $u$ acquires the values of $z(u)$ that satisfy the condition $z\in \mathbb{R}$. From Fig. \ref{fig2}, it can be deduced that by reducing $\varphi$, the elliptical wormhole becomes similar to the Beltrami wormhole in Fig. \ref{fig3}.

As shown in Fig. \ref{fig3}, by increasing $b$, the wormhole's throat becomes shorter.

\begin{figure}[H]
\centering
\subfigure[With $d=0.7$, $r=1$ and $-\frac{\pi}{2}\le \frac{u}{r} \le \frac{\pi}{2}$.]{%
\label{fig41}%
\includegraphics[height=6cm,width=7cm]{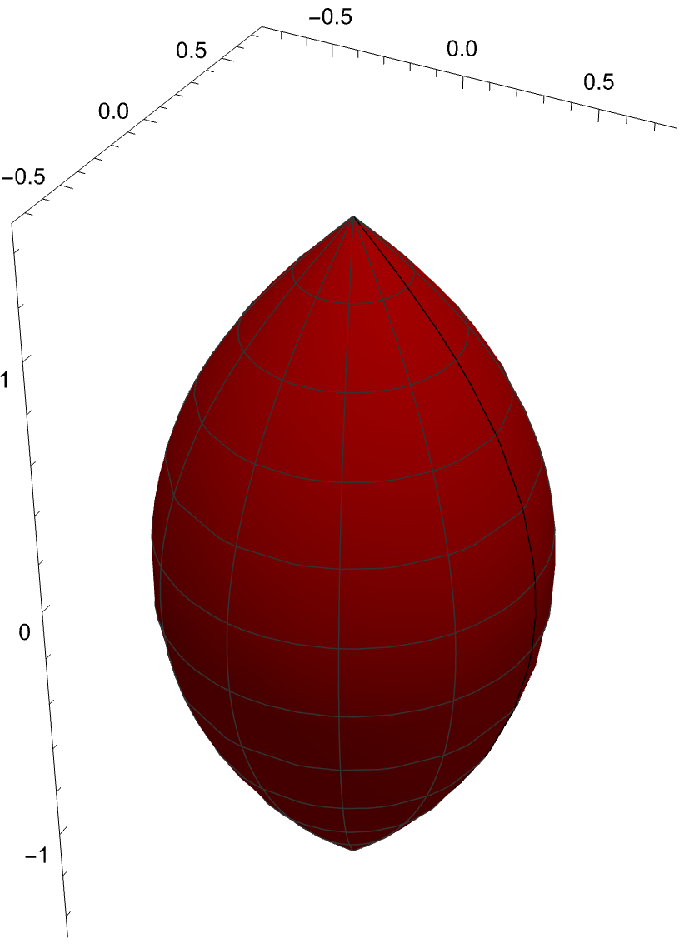}}%
\qquad
\subfigure[With $d=2$, $r=2.5$ and $-\frac{\pi}{6}\le \frac{u}{r} \le \frac{\pi}{6}$.]{%
\label{fig42}%
\includegraphics[height=6cm]{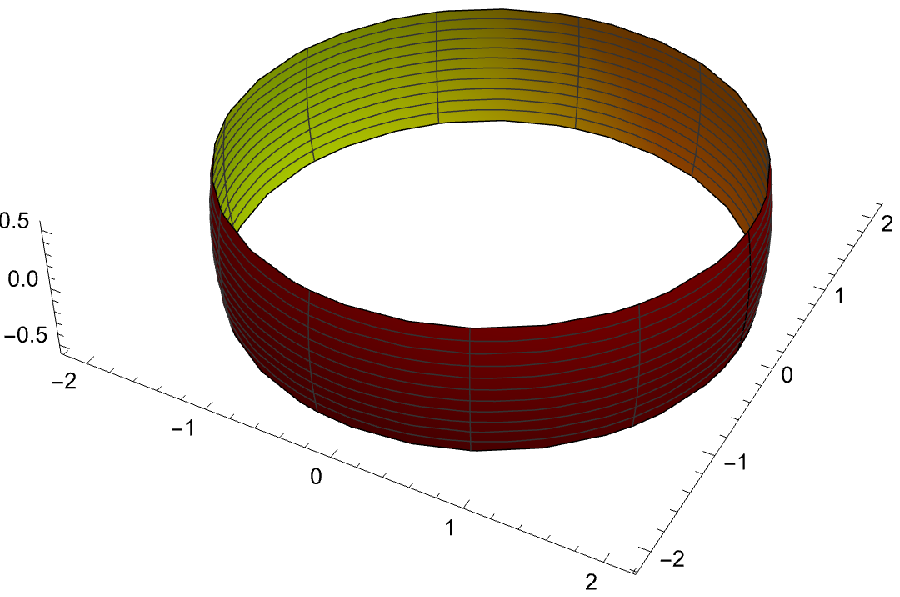}}%
\caption{Two distinct surfaces related to $R(u)=d\, \mathrm{cos}\left(\frac{u}{r}\right)$ identified as a particular solution of Eq. \eqref{constGausscurvature} when $k>0$ are described on two different cases $d<1$ and $d>1$.  }\label{fig4}. 
\end{figure}

Given Eq. \eqref{plusK} and considering $\phi=0$, it is clear that Eq. \eqref{plusK} is reduced to the form $R(u)=d\, \mathrm{cos}\left(\frac{u}{r}\right)$ so that assuming $d<1$, $d>1$ and $d=1$ gives rise to three distinct surfaces, the two first surfaces are   shown in Fig. \ref{fig4}, and the third surface is related to the ordinary sphere with the radius $r$.

Figure \ref{fig4} is plotted by considering $x(u,v) = d\, \mathrm{cos}\left(\frac{u}{r}\right) \mathrm{cos} (v)$, $y(u,v) = d\, \mathrm{cos}\left(\frac{u}{r}\right) \mathrm{sin}(v)$ and $z(u) =r\, E(\frac{u}{r},\frac{d^2}{r^2})$ with $\frac{d^2}{r^2}<1$, and $v\in [0,2\pi]$. The only restriction for sectors $d<1$ and $d>1$ is that
the integrand $\sqrt{1-(R'(u))^2}$ is always real \cite{Eisenhart-Spivak}.

%%%%%%%%%%%%%%%%%%%%%%%%%%%%%%%%%%%%%%%%%%%%%%%%

%%% SEC 4

%%%%%%%%%%%%%%%%%%%%%%%%%%%%%%%%%%%%%%%%%%%%%%%%

\section{Hyperbolic wormhole} \label{sec4}

In this section, we solve Eq. \eqref{DiracEq4} in a curved space produced by the hyperbolic wormhole  described by a constant negative Gaussian curvature, $\mathcal{K}<0$, by considering  the meridian-hyperbolic function $R(u) = b_{2}\, \mathrm{cosh}\left(\frac{u}{r}\right)$, with the wormhole's radius $b_{2}$ at the wormhole's throat, where $u=0$, and the radius of curvature of the constant negative Gaussian curvature (the wormhole surface) $r$ along the $u$ direction. 
Then, if we change to the new variable $\mathcal{X} = \frac{r}{b_{2}} R'(u) \equiv\mathrm{sinh}(\frac{u}{r})$ in Eq. \eqref{DiracEq4}, we find
\begin{equation}\label{DiracEq5}
\begin{split}	
&\left(1+\mathcal{X}^2\right)\psi_{1}''(\mathcal{X})+2(\tau+1)\mathcal{X}\psi_{1}'(\mathcal{X})+\left[k^2r^2+\frac{\frac{mr}{b_{2}}\mathcal{X}-\frac{m^2r^2}{b_{2}^2}}{1+\mathcal{X}^2}\right.\\
&\left.+\left\{\tau^2-\frac14\right\}\frac{\mathcal{X}^2}{1+\mathcal{X}^2}
+\left(\tau+\frac12\right)
\right]\psi_{1}(\mathcal{X}) = 0,  
\end{split}
\end{equation}
in which we suppose that $\tau^2-1/4=0$. This implies that $\tau=-1/2$ and $1/2$ and the analytical solutions for the Jacobi differential equation,
\begin{equation}\label{JacobiDiffEq}
(1-x^2)Y''(x)+\left[B-A-(B+A+2)x\right]Y'(x)+n\left[n+B+A+1\right]Y(x)=0.
\end{equation}
Thus, in the following subsections, we provide exact solutions of Eq. \eqref{DiracEq5} to two possible scenarios $\tau=-1/2$ and $1/2$.

%%%%% SUBSEC 4.1

\subsection{Exact solutions with $\tau=-1/2$}
With $\tau=-1/2$, Eq. \eqref{DiracEq5} becomes
\begin{equation}\label{JacobiDiffEq1}
\begin{split}	
\left(1+\mathcal{X}^2\right)\psi_{1}''(\mathcal{X})+\mathcal{X}\psi_{1}'(\mathcal{X})+\left[k^2r^2+\frac{\frac{mr}{b_{2}}\mathcal{X}-\frac{m^2r^2}{b_{2}^2}}{1+\mathcal{X}^2}\right]\psi_{1}(\mathcal{X}) = 0.
\end{split}
\end{equation}
If we substitute the ansatz solution
\begin{equation}\label{ansatz}
\psi_{1}\left(\mathcal{X}\right) = \left(1+\mathrm{i}\mathcal{X}\right)^{\bar{\alpha}}\left(1-\mathrm{i}\mathcal{X}\right)^{\bar{\beta}}\mathcal{Q}\left(\mathcal{X}\right)
\end{equation} 
into Eq. \eqref{JacobiDiffEq1}, we find
\begin{equation}\label{JacobiDiffEq11}
\begin{split}	
&\left(1+\mathcal{X}^2\right)\mathcal{Q}''(\mathcal{X})+2\left[\left(\bar{\alpha}+\bar{\beta}+\frac12\right)\mathcal{X}+\mathrm{i}\left(\bar{\alpha}-\bar{\beta}\right)\right]\mathcal{Q}'(\mathcal{X})\\
&+\left[k^2r^2+\left(\bar{\alpha}+\bar{\beta}\right)^{2}\right]\mathcal{Q}(\mathcal{X}) = 0.
\end{split}
\end{equation}
From Eqs. \eqref{JacobiDiffEq1} and \eqref{JacobiDiffEq11}, we see that
\begin{subequations}\label{finding2Eq}
\begin{align}
&\frac{mr}{b}-\mathrm{i}\bar{\alpha}+2\mathrm{i}\bar{\alpha}^{2}+\mathrm{i}\bar{\beta}-2\mathrm{i}\bar{\beta}^{2}=0,\\
-&\frac{m^2r^2}{b^2}-2\left(\bar{\alpha}^{2}+\bar{\beta}^{2}\right)+\bar{\alpha}+\bar{\beta}=0.
\end{align}
\end{subequations}
Thereby, the ansatz parameters $\bar{\alpha}_{\jmath}$ and $\bar{\beta}_{\jmath}$ are found to be
\begin{subequations}\label{finding2Eq-1}
\begin{align}
& \bar{\alpha}_{1} = -\frac{\mathrm{i}mr}{2b_{2}}, \qquad \bar{\beta}_{1} = \frac{\mathrm{i}mr}{2b_{2}},\\
& \bar{\alpha}_{2} = -\frac{\mathrm{i}mr}{2b_{2}}, \qquad \bar{\beta}_{2} = \frac{b_{2}-\mathrm{i}mr}{2b_{2}},\\
& \bar{\alpha}_{3} = \frac{b_{2}+\mathrm{i}mr}{2b_{2}}, \qquad \bar{\beta}_{3} = \frac{\mathrm{i}mr}{2b_{2}},\\
& \bar{\alpha}_{4} = \frac{b_{2}+\mathrm{i}mr}{2b_{2}}, \qquad \bar{\beta}_{4} = \frac{b_{2}-\mathrm{i}mr}{2b_{2}}.
\end{align}
\end{subequations}
We obtain a wave equation similar to Eq. \eqref{JacobiDiffEq} through the change of variable $\mathcal{X} = -\mathrm{i}\mathcal{S}$ in Eq. \eqref{JacobiDiffEq11}:
\begin{equation}\label{JacobiDiffEq2}
\begin{split}	
&\left(1-\mathcal{S}^2\right)\mathcal{Q}''(\mathcal{S})+2\left[\left(\bar{\alpha}_{\jmath}-\bar{\beta}_{\jmath}\right)-\left(\bar{\alpha}_{\jmath}+\bar{\beta}_{\jmath}+\frac12\right)\mathcal{S}\right]\mathcal{Q}'(\mathcal{S})\\
&-\left[k^2r^2+\left(\bar{\alpha}_{\jmath}+\bar{\beta}_{\jmath}\right)^{2}\right]\mathcal{Q}(\mathcal{S}) = 0.
\end{split}
\end{equation}
If we compare Eqs. \eqref{JacobiDiffEq} and \eqref{JacobiDiffEq2}, we find that the Jacobi equation parameters $A_{\jmath}$ and $B_{\jmath}$ can be written as follows
\begin{equation}\label{JacobiParameters}
A_{\jmath}=-\frac12+2\bar{\beta}_{\jmath}, \qquad B_{\jmath}=-\frac12+2\bar{\alpha}_{\jmath},
\end{equation}
and the energy levels in terms of $\bar{\alpha}_{\jmath}$ and $\bar{\beta}_{\jmath}$ indicated by Eq. \eqref{finding2Eq-1} are generally given by
\begin{equation}\label{energyLevels1}
E_{nm\jmath}^{(HWH,\tau=-1/2)} =\pm\frac{\mathrm{i}}{r}\sqrt{n^{2}-M^{2}r^2+2n(\bar{\alpha}_{\jmath}+\bar{\beta}_{\jmath})+(\bar{\alpha}_{\jmath}+\bar{\beta}_{\jmath})^{2}}, 
\end{equation}
where $n=1$, 2, 3, ..., and the values of $m$ are given after Eq. \eqref{PrimWF}. The corresponding eigenfunctions are expressed in terms of the hypergeometric function ${}_{2}F_{1}$ as
\begin{equation}\label{WFJacobiEq}
\begin{split}
\mathcal{Q}_{nm\jmath}^{(HWH,\tau=-1/2)}(\mathcal{S}) &= \mathcal{C}_{1}\,\,\,{}_{2}F_{1}\left(-n,n+2(\bar{\alpha}_{\jmath}+\bar{\beta}_{\jmath}),\frac12+2\bar{\beta}_{\jmath},\frac12(\mathcal{S}-1)\right)\\
&+2^{-\frac12+2\bar{\beta}_{\jmath}}\left(\mathcal{S}-1\right)^{\frac12-2\bar{\beta}_{\jmath}}\mathcal{C}_{2}\,\,\,{}_{2}F_{1}\left(\frac12+n+2\bar{\alpha}_{\jmath},\frac12-n-2\bar{\beta}_{\jmath},\frac32-2\bar{\beta}_{\jmath},\frac12(1-\mathcal{S})\right).
\end{split}
\end{equation}

In Table \ref{tabelWHWcase1}, we display the energy levels, Eq \eqref{energyLevels1}, and wave-functions, Eq. \eqref{WFJacobiEq}, for the values of the parameters defined in Eq. \eqref{finding2Eq-1} for each of $\jmath=1$, 2, 3 and 4.   As in Ref. \cite{RojjanasonEPJC1920} (2019 paper), the energy levels are complex values, which suggest dissipation of energy; that is, unstable states. Thus we can make an observation analogous to that paper, Ref. \cite{RojjanasonEPJC1920}, which investigated charged fermions, whereas we consider neutral fermions in wormholes; in a similar way, we can interpret the states of those fermions are quasinormal modes that would be decaying for negative imaginary frequencies and unstable is the frequencies are positive imaginary.

The simplest, and mutually quite similar, expressions of the energies are with  $\jmath=1$ and 4.  
 For all the classes $\jmath$, the energy scales with $\frac 1r$, so that smaller values of $r$ lead to larger energies.

%%%%%%%%%%%%%%%
%%%%Table%%%%%%
%%%%%%%%%%%%%%%

\begin{table}[H]
\caption{ Classes of solutions of Eq. \eqref{JacobiDiffEq2} for different values of $\bar{\alpha}_{\jmath}$ and $\bar{\beta}_{\jmath}$ in Eq. \eqref{finding2Eq-1}} % title name of the table
\centering % centering table
\begin{tabular}{c c c } % creating 10 columns
\hline\hline % inserting double-line
$\bar{\alpha}_{\jmath}$ & $\bar{\beta}_{\jmath}$ & Solutions 
\\
\hline % inserts single-line % Entering 1st row
& & $E_{nm1}^{(HWH,\tau=-1/2)}= \pm \frac{\mathrm{i}}{r}\sqrt{n^{2}-M^2 r^2}$ \\[0.5ex] \cline{3-3}
\raisebox{1.5ex}{$-\frac{\mathrm{i}mr}{2b_{2}}$} & \raisebox{1.5ex}{$\frac{\mathrm{i}mr}{2b_{2}}$} 
&$\mathcal{Q}_{nm1}^{(HWH,\tau=-1/2)}=\mathcal{C}_{1}\,\,\,{}_{2}F_{1}\left(-n,n,\frac12+\frac{\mathrm{i}mr}{b_{2}},\frac12(\mathcal{S}-1)\right)$\\[1ex]
& & $+2^{-\frac12+\frac{\mathrm{i}mr}{b_{2}}}\left(\mathcal{S}-1\right)^{\frac12-\frac{\mathrm{i}mr}{b_{2}}}\mathcal{C}_{2}\,\,\,{}_{2}F_{1}\left(\frac12-n-\frac{\mathrm{i}mr}{b_{2}},\frac12+n-\frac{\mathrm{i}mr}{b_{2}},\frac32-\frac{\mathrm{i}mr}{b_{2}},\frac12(1-\mathcal{S})\right)
  $ \\[1ex]% Entering 2nd row
\hline
& &$E_{nm2}^{(HWH,\tau=-1/2)}=\pm\frac{\mathrm{i}}{2r}\sqrt{(1+2n)^2-4M^2r^2-\frac{4mr}{b_{2}}\left(\mathrm{i}(1+2n)+\frac{m r}{b_{2}}\right)}$ \\[1ex] \cline{3-3}
\raisebox{1.5ex}{$-\frac{\mathrm{i}mr}{2b_{2}}$} & \raisebox{1.5ex}{$\frac{b_{2}-\mathrm{i}mr}{2b_{2}}$}
&$\mathcal{Q}_{nm2}^{(HWH,\tau=-1/2)}=\mathcal{C}_{1}\,\,\,{}_{2}F_{1}\left(-n,1+n-\frac{2\mathrm{i}mr}{b_{2}},\frac32-\frac{\mathrm{i}mr}{b_{2}},\frac12(\mathcal{S}-1)\right)$\\[1ex]
& & $+2^{\frac12-\frac{\mathrm{i}mr}{b_{2}}}\left(\mathcal{S}-1\right)^{-\frac12+\frac{\mathrm{i}mr}{b_{2}}}\mathcal{C}_{2}\,\,\,{}_{2}F_{1}\left(\frac12+n-\frac{\mathrm{i}mr}{b_{2}},-\frac12-n+\frac{\mathrm{i}mr}{b_{2}},\frac12+\frac{\mathrm{i}mr}{b_{2}},\frac12(1-\mathcal{S})\right)
$ \\[1ex]% Entering 3rd row
\hline  
& &$E_{nm3}^{(HWH,\tau=-1/2)}=\pm\frac{\mathrm{i}}{2r}\sqrt{(1+2n)^2-4M^2r^2+\frac{4mr}{b_{2}}\left(\mathrm{i}(1+2n)-\frac{m r}{b_{2}}\right)}$ \\[1ex] \cline{3-3}
\raisebox{1.5ex}{$\frac{b_{2}+\mathrm{i}mr}{2b_{2}}$} &
\raisebox{1.5ex}{$\frac{\mathrm{i}mr}{2b_{2}}$}
&$\mathcal{Q}_{nm3}^{(HWH,\tau=-1/2)}=\mathcal{C}_{1}\,\,\,{}_{2}F_{1}\left(-n,1+n+\frac{2\mathrm{i}mr}{b_{2}},\frac12+\frac{\mathrm{i}mr}{b_{2}},\frac12(\mathcal{S}-1)\right)$\\[1ex]
& & $+2^{-\frac12+\frac{\mathrm{i}mr}{b_{2}}}\left(\mathcal{S}-1\right)^{\frac12-\frac{\mathrm{i}mr}{b_{2}}}\mathcal{C}_{2}\,\,\,{}_{2}F_{1}\left(\frac12-n-\frac{\mathrm{i}mr}{b_{2}},\frac32+n+\frac{\mathrm{i}mr}{b_{2}},\frac32-\frac{\mathrm{i}mr}{b_{2}},\frac12(1-\mathcal{S})\right)
$ \\[1ex]% [1ex] adds vertical space
\hline
& &$E_{nm4}^{(HWH,\tau=-1/2)}= \pm \frac{\mathrm{i}}{r}\sqrt{(1+n)^{2}-M^2 r^2}$ \\[1ex] \cline{3-3}
\raisebox{1.5ex}{$\frac{b_{2}+\mathrm{i}mr}{2b_{2}}$} &
\raisebox{1.5ex}{$\frac{b_{2}-\mathrm{i}mr}{2b_{2}}$}
&$\mathcal{Q}_{nm4}^{(HWH,\tau=-1/2)}=\mathcal{C}_{1}\,\,\,{}_{2}F_{1}\left(-n,2+n,\frac32-\frac{\mathrm{i}mr}{b_{2}},\frac12(\mathcal{S}-1)\right)$\\[1ex]
& & $+2^{\frac12-\frac{\mathrm{i}mr}{b_{2}}}\left(\mathcal{S}-1\right)^{-\frac12+\frac{\mathrm{i}mr}{b_{2}}}\mathcal{C}_{2}\,\,\,{}_{2}F_{1}\left(-\frac12-n+\frac{\mathrm{i}mr}{b_{2}},\frac32+n+\frac{\mathrm{i}mr}{b_{2}},\frac12+\frac{\mathrm{i}mr}{b_{2}},\frac12(1-\mathcal{S})\right)
$ \\[1ex]% [1ex] adds vertical space

\hline
\hline % inserts single-line
\end{tabular}
\label{tabelWHWcase1}
\end{table}

%%%%%%
%  SUBSEC 4.2
%%%%%%

\subsection{Exact solutions with $\tau=1/2$}

Now let us turn to $\tau=1/2$. If we substitute Eq. \eqref{ansatz} into Eq. \eqref{DiracEq5} with $\tau=1/2$, we find 
\begin{equation}\label{JacobiDiffEq3}
\begin{split}	
&\left(1+\mathcal{X}^2\right)\mathcal{Q}''(\mathcal{X})+2\left[\left(\bar{\alpha}+\bar{\beta}+\frac32\right)\mathcal{X}+\mathrm{i}\left(\bar{\alpha}-\bar{\beta}\right)\right]\mathcal{Q}'(\mathcal{X})\\
&+\left[k^2r^2+1+\left(\bar{\alpha}+\bar{\beta}\right)\left(2+\bar{\alpha}+\bar{\beta}\right)\right]\mathcal{Q}(\mathcal{X}) = 0.
\end{split}
\end{equation}
As in the previous section, we find
\begin{subequations}\label{finding2Eq2}
\begin{align}
&\frac{mr}{b}+\mathrm{i}\bar{\alpha}+2\mathrm{i}\bar{\alpha}^{2}-\mathrm{i}\bar{\beta}-2\mathrm{i}\bar{\beta}^{2}=0,\\
-&\frac{m^2r^2}{b^2}-2\left(\bar{\alpha}^{2}+\bar{\beta}^{2}\right)-\bar{\alpha}-\bar{\beta}=0,
\end{align}
\end{subequations}
and the relevant ansatz parameters $\bar{\alpha}_{j}$ and $\bar{\beta}_{j}$ are  thus given by
\begin{subequations}\label{finding2Eq2-1}
	\begin{align}
		& \bar{\alpha}_{1} = \frac{\mathrm{i}mr}{2b_{2}}, \qquad \bar{\beta}_{1} = -\frac{\mathrm{i}mr}{2b_{2}},\\
		& \bar{\alpha}_{2} = \frac{\mathrm{i}mr}{2b_{2}}, \qquad \bar{\beta}_{2} = -\frac{b_{2}-\mathrm{i}mr}{2b_{2}},\\
		& \bar{\alpha}_{3} = -\frac{b_{2}+\mathrm{i}mr}{2b_{2}}, \qquad \bar{\beta}_{3} = -\frac{\mathrm{i}mr}{2b_{2}},\\
		& \bar{\alpha}_{4} = -\frac{b_{2}+\mathrm{i}mr}{2b_{2}}, \qquad \bar{\beta}_{4} = -\frac{b_{2}-\mathrm{i}mr}{2b_{2}}.
	\end{align}
\end{subequations}
We find a wave equation analogous to Eq. \eqref{JacobiDiffEq} by the change of variable $\mathcal{X} = -\mathrm{i}\mathcal{Z}$ in Eq. \eqref{JacobiDiffEq3}:
\begin{equation}\label{JacobiDiffEq4}
	\begin{split}	
		&\left(1-\mathcal{Z}^2\right)\mathcal{Q}''(\mathcal{Z})+2\left[\left(\bar{\alpha}_{j}-\bar{\beta}_{j}\right)-\left(\bar{\alpha}_{j}+\bar{\beta}_{j}+\frac32\right)\mathcal{Z}\right]\mathcal{Q}'(\mathcal{Z})\\
		&-\left[k^2r^2+1+\left(\bar{\alpha}_{j}+\bar{\beta}_{j}\right)\left(\bar{\alpha}_{j}+\bar{\beta}_{j}+2\right)\right]\mathcal{Q}(\mathcal{Z}) = 0.
	\end{split}
\end{equation}
Now, by comparing Eqs. \eqref{JacobiDiffEq} and \eqref{JacobiDiffEq2}, the Jacobi equation parameters $A$ and $B$ can be written as follows,
\begin{equation}\label{JacobiParameters}
	A=\frac12+2\bar{\beta}_{j}, \qquad B=\frac12+2\bar{\alpha}_{j},
\end{equation}
and the energy levels with regard to Eq. \eqref{finding2Eq2-1} are generally given by
\begin{equation}\label{energyLevels}
E_{nmj}^{(HWH,\tau=1/2)} =\pm\frac{\mathrm{i}}{r}\sqrt{n^{2}-M^{2}r^2+2n(1+\bar{\alpha}_{j}+\bar{\beta}_{j})+(1+\bar{\alpha}_{j}+\bar{\beta}_{j})^{2}}, 
\end{equation}
and the corresponding eigenfunctions can be written in terms of the hypergeometric function,
\begin{equation}\label{WFJacobiEq2}
\begin{split}
\mathcal{Q}_{nmj}^{(HWH,\tau=1/2)}(\mathcal{Z}) &= \mathcal{C}_{1}\,\,\,{}_{2}F_{1}\left(-n,n+2(1+\bar{\alpha}_{j}+\bar{\beta}_{j}),\frac32+2\bar{\beta}_{j},\frac12(\mathcal{Z}-1)\right)\\
&+2^{\frac12+2\bar{\beta}_{j}}\left(\mathcal{Z}-1\right)^{-\frac12-2\bar{\beta}_{j}}\mathcal{C}_{2}\,\,\,{}_{2}F_{1}\left(\frac32+n+2\bar{\alpha}_{j},-\frac12-n-2\bar{\beta}_{j},\frac12-2\bar{\beta}_{j},\frac12(1-\mathcal{Z})\right).
\end{split}
\end{equation}
We can recap the present scenario's energy levels and corresponding wave functions in terms of the solutions in  Table \ref{tabelWHWcase1} via the correspondences:
\begin{subequations}\label{Solutionscase2}
\begin{align}
& E_{nm1}^{(HWH,\tau=1/2)} \equiv E_{nm4}^{(HWH,\tau=-1/2)}, \qquad \mathcal{Q}_{nm1}^{(HWH,\tau=1/2)} \equiv \mathcal{Q}_{nm4}^{(HWH,\tau=-1/2)},\\
&E_{nm2}^{(HWH,\tau=1/2)} \equiv E_{nm3}^{(HWH,\tau=-1/2)}, \qquad \mathcal{Q}_{nm2}^{(HWH,\tau=1/2)} \equiv \mathcal{Q}_{nm3}^{(HWH,\tau=-1/2)},\\
&E_{nm3}^{(HWH,\tau=1/2)} \equiv E_{nm2}^{(HWH,\tau=-1/2)}, \qquad \mathcal{Q}_{nm3}^{(HWH,\tau=1/2)} \equiv \mathcal{Q}_{nm2}^{(HWH,\tau=-1/2)},\\
&E_{nm4}^{(HWH,\tau=1/2)} \equiv E_{nm1}^{(HWH,\tau=-1/2)}, \qquad \mathcal{Q}_{nm4}^{(HWH,\tau=1/2)} \equiv \mathcal{Q}_{nm1}^{(HWH,\tau=-1/2)}.
\end{align}
\end{subequations}

%%%%%%%%%%%%%%%%%%%%%%%%%%%%%%%%%%%%%%%%%%%%%%%%

%%% SEC 5

%%%%%%%%%%%%%%%%%%%%%%%%%%%%%%%%%%%%%%%%%%%%%%%%

\section{Elliptic wormhole \label{sec5}}

In this section, we consider $\mathcal{K}<0$, that is, the first term in Eq. \eqref{minusK} describing a kind of surface of revolution recognized as the elliptic wormhole.
Then, in order to solve Eq. \eqref{DiracEq4} in the presence of the elliptic wormhole, we set the relevant function $R(u) = b_{1}\, \mathrm{sinh}\left(\frac{u}{r}\right)$ in Eq. \eqref{DiracEq4} and apply the change of variable $\chi = \frac{r}{b_{1}} R'(u) \equiv\mathrm{cosh}(\frac{u}{r})$. Thus, we get to
\begin{comment}
\begin{equation}
\begin{split}
&\left(1-\mathcal{\chi}^2\right)\psi_{1}''(\mathcal{\chi})+2\tau\mathcal{\chi}\psi_{1}'(\mathcal{\chi})-\left[k^2r^2+\frac{\frac{mr}{b_{1}}\mathcal{\chi}-\frac{m^2r^2}{b_{1}^2}}{\mathcal{\chi}^2-1}\right.\nonumber\\
&\left.+\left\{\left(\tau+1\right)^2-\frac14\right\}\frac{\mathcal{\chi}^2}{\mathcal{\chi}^2-1}
-\left(\tau+\frac12\right)
\right]\psi_{1}(\mathcal{\chi}) = 0,
\end{split}
\end{equation}
\end{comment}
\begin{subequations}\label{DiracEq6}
\begin{align}	
&\left(1-\chi^2\right)\psi_{1}''(\chi)-\chi\psi_{1}'(\chi)-\left[k^2r^2+\frac{\frac{mr}{b_{1}}\chi-\frac{m^2r^2}{b_{1}^2}}{\chi^{2}-1}\right]\psi_{1}(\chi) = 0, \quad \text{for} \quad \tau=-\frac12, \label{EWHEq1}\\
&\left(1-\chi^{2}\right)\psi_{1}''(\chi)-3\chi\psi_{1}'(\chi)-\left[k^2r^2+\frac{\frac{mr}{b_{1}}\chi-\frac{m^2r^2}{b_{1}^2}}{\chi^{2}-1}+1\right]\psi_{1}(\chi) = 0, \quad \text{for} \quad \tau=\frac12, \label{EWHEq2}
\end{align}
\end{subequations}
where $0<b_{1}=r\mathrm{cos}\varphi<r$. 
In this way, with the ansatz solution 
\begin{equation}\label{ansatz2}
\psi_{1}\left(\chi\right) = \left(1+\chi\right)^{\Bar{\Bar{\alpha}}}\left(1-\chi\right)^{\Bar{\Bar{\beta}}}\mathcal{P}\left(\chi\right),
\end{equation}
and the same analytical method as in the previous section, we are able to give the solutions of Eqs. \eqref{EWHEq1} and \eqref{EWHEq2}. First, we present the solutions of Eq. \eqref{EWHEq1} through the table below
%%%%%%%%%%%%%%%
%%%%Table%%%%%%
%%%%%%%%%%%%%%%
\begin{table}[H]
\caption{Classifying the solutions of Eq. \eqref{EWHEq1} in terms of different values of $\Bar{\Bar{\alpha}}_{\jmath}$ and $\Bar{\Bar{\beta}}_{\jmath}$ } % title name of the table
\centering % centering table
\begin{tabular}{c c c } % creating 10 columns
\hline\hline % inserting double-line
$\Bar{\Bar{\alpha}}_{\jmath}$ & $\Bar{\Bar{\beta}}_{\jmath}$ & Solutions 
\\
\hline % inserts single-line % Entering 1st row
& & $E_{nm1}^{(EWH,\tau=-1/2)}= \pm \frac{\mathrm{i}}{r}\sqrt{n^{2}-M^2 r^2}$ \\[0.5ex] \cline{3-3}
\raisebox{1.5ex}{$-\frac{mr}{2b_{1}}$} & \raisebox{1.5ex}{$\frac{mr}{2b_{1}}$} 
&$\mathcal{P}_{nm1}^{(EWH,\tau=-1/2)}=\mathcal{C}_{1}\,\,\,{}_{2}F_{1}\left(-n,n,\frac12+\frac{mr}{b_{1}},\frac12(\mathcal{\chi}-1)\right)$\\[1ex]
& & $+2^{-\frac12+\frac{mr}{b_{1}}}\left(\chi-1\right)^{\frac12-\frac{mr}{b_{1}}}\mathcal{C}_{2}\,\,\,{}_{2}F_{1}\left(\frac12-n-\frac{mr}{b_{1}},\frac12+n-\frac{mr}{b_{1}},\frac32-\frac{mr}{b_{1}},\frac12(1-\chi)\right)
$ \\[1ex]% Entering 2nd row
\hline
& &$E_{nm2}^{(EWH,\tau=-1/2)}=\pm\frac{\mathrm{i}}{2r}\sqrt{(1+2n)^2-4M^2r^2-\frac{4mr}{b_{1}}\left((1+2n)-\frac{m r}{b_{1}}\right)}$ \\[1ex] \cline{3-3}
\raisebox{1.5ex}{$-\frac{mr}{2b_{1}}$} & \raisebox{1.5ex}{$\frac{b_{1}-mr}{2b_{1}}$}
&$\mathcal{P}_{nm2}^{(EWH,\tau=-1/2)}=\mathcal{C}_{1}\,\,\,{}_{2}F_{1}\left(-n,1+n-\frac{2mr}{b_{1}},\frac32-\frac{mr}{b_{1}},\frac12(\chi-1)\right)$\\[1ex]
& & $+2^{\frac12-\frac{mr}{b_{1}}}\left(\chi-1\right)^{-\frac12+\frac{mr}{b_{1}}}\mathcal{C}_{2}\,\,\,{}_{2}F_{1}\left(\frac12+n-\frac{mr}{b_{1}},-\frac12-n+\frac{mr}{b_{1}},\frac12+\frac{mr}{b_{1}},\frac12(1-\chi)\right)
$ \\[1ex]% Entering 3rd row
\hline  
& &$E_{nm3}^{(EWH,\tau=-1/2)}=\pm\frac{\mathrm{i}}{2r}\sqrt{(1+2n)^2-4M^2r^2+\frac{4mr}{b_{1}}\left((1+2n)+\frac{m r}{b_{1}}\right)}$ \\[1ex] \cline{3-3}
\raisebox{1.5ex}{$\frac{b_{1}+mr}{2b_{1}}$} &
\raisebox{1.5ex}{$\frac{mr}{2b_{1}}$}
&$\mathcal{P}_{nm3}^{(EWH,\tau=-1/2)}=\mathcal{C}_{1}\,\,\,{}_{2}F_{1}\left(-n,1+n+\frac{2mr}{b_{1}},\frac12+\frac{mr}{b_{1}},\frac12(\chi-1)\right)$\\[1ex]
& & $+2^{-\frac12+\frac{mr}{b_{1}}}\left(\chi-1\right)^{\frac12-\frac{mr}{b_{1}}}\mathcal{C}_{2}\,\,\,{}_{2}F_{1}\left(\frac12-n-\frac{mr}{b_{1}},\frac32+n+\frac{mr}{b_{1}},\frac32-\frac{mr}{b_{1}},\frac12(1-\chi)\right)
$ \\[1ex]% [1ex] adds vertical space
\hline
& &$E_{nm4}^{(EWH,\tau=-1/2)}= \pm \frac{\mathrm{i}}{r}\sqrt{(1+n)^{2}-M^2 r^2}$ \\[1ex] \cline{3-3}
\raisebox{1.5ex}{$\frac{b_{1}+mr}{2b_{1}}$} &
\raisebox{1.5ex}{$\frac{b_{1}-mr}{2b_{1}}$}
&$\mathcal{P}_{nm4}^{(EWH,\tau=-1/2)}=\mathcal{C}_{1}\,\,\,{}_{2}F_{1}\left(-n,2+n,\frac32-\frac{mr}{b_{1}},\frac12(\chi-1)\right)$\\[1ex]
& & $+2^{\frac12-\frac{mr}{b_{1}}}\left(\chi-1\right)^{-\frac12+\frac{mr}{b_{1}}}\mathcal{C}_{2}\,\,\,{}_{2}F_{1}\left(-\frac12-n+\frac{mr}{b_{1}},\frac32+n+\frac{mr}{b_{1}},\frac12+\frac{mr}{b_{1}},\frac12(1-\chi)\right)
$ \\[1ex]% [1ex] adds vertical space
		
\hline
\hline % inserts single-line
\end{tabular}
\label{tabelEHWcase1}
\end{table}
We can then summarize the solutions of Eq. \eqref{EWHEq2} according to the solutions presented in Table \ref{tabelEHWcase1} as follows

\begin{subequations}\label{SolutionsEWHcase2}
\begin{align}
& E_{nm1}^{(EWH,\tau=1/2)} \equiv E_{nm4}^{(EWH,\tau=-1/2)}, \qquad \mathcal{P}_{nm1}^{(EWH,\tau=1/2)} \equiv \mathcal{P}_{nm4}^{(EWH,\tau=-1/2)},\\
&E_{nm2}^{(EWH,\tau=1/2)} \equiv E_{nm3}^{(EWH,\tau=-1/2)}, \qquad \mathcal{P}_{nm2}^{(EWH,\tau=1/2)} \equiv \mathcal{P}_{nm3}^{(EWH,\tau=-1/2)},\\
&E_{nm3}^{(EWH,\tau=1/2)} \equiv E_{nm2}^{(EWH,\tau=-1/2)}, \qquad \mathcal{P}_{nm3}^{(EWH,\tau=1/2)} \equiv \mathcal{P}_{nm2}^{(EWH,\tau=-1/2)},\\
&E_{nm4}^{(EWH,\tau=1/2)} \equiv E_{nm1}^{(EWH,\tau=-1/2)}, \qquad \mathcal{P}_{nm4}^{(EWH,\tau=1/2)} \equiv \mathcal{P}_{nm1}^{(EWH,\tau=-1/2)}.
\end{align}
\end{subequations}

%%%%%%%%%%%%%%%%%%%%%%%%%%%%%%%%%%%%%%%%%%%%%%%%

%%% SEC 6

%%%%%%%%%%%%%%%%%%%%%%%%%%%%%%%%%%%%%%%%%%%%%%%%

\section{Beltrami wormhole \label{sec6}}

 Finally, in this section, we solve the wave equation \eqref{DiracEq4} for a curved $(1+2)$-dimensional space-time produced by the Beltrami wormhole, which  
corresponds to $\mathcal{K}<0$. Hence, substituting $R(u) = b\, e^{\frac{u}{r}}$
in Eq. \eqref{DiracEq4}, we arrive at
\begin{equation}\label{DiracEq7}	
\psi_{1}''(u)+\frac{(2\tau+1)}{r}\psi_{1}'(u)+\left[k^2+\frac{(\tau+1/2)^{2}}{r^2}+\frac{e^{-\frac{2u}{r}}m\left(b e^{\frac{u}{r}}-mr\right)}{b^{2}r}\right]\psi_{1}(u) = 0
\end{equation}
\begin{comment}
where the general solutions of Eqs. \eqref{EWHEq3} and \eqref{EWHEq4} are respectively given by the following expression
\begin{equation}
\begin{split}
\psi_{1}^{(BWH,\tau=-1/2)}(u) &= e^{-\frac{mr}{b}e^{-\frac{u}{r}}}\left(e^{-\frac{u}{r}}\right)^{\mathrm{i} k r}\left[\mathcal{C}_{1}\,U\left(\mathrm{i} k r,1+2\mathrm{i} k r, \frac{2mr}{b}e^{-\frac{u}{r}}\right)\right.\\
&\left.+\mathcal{C}_{2}\, \Gamma_{-\mathrm{i}k r}^{2\mathrm{i}k r} \left(\frac{2mr}{b}e^{-\frac{u}{r}}\right)\right], \quad \text{for} \quad \tau=-\frac{1}{2}\\
\psi_{1}^{(BWH,\tau=-3/2)}(u) &= e^{-\frac{u}{r}}\,\, \psi_{1}^{(BWH,\tau=-1/2)}(u), \quad \text{for} \quad \tau=-\frac{3}{2} 
\end{split}
\end{equation}
\end{comment}
 Hereafter, we solve Eq. \eqref{DiracEq7} with the Nikiforov-Uvarov (NU) method, such that the solutions of Schr\"{o}dinger-like second-order differential equations Ref. \cite{nu}
 %(see Refs. \cite{nu} and Section 3 of Ref. \cite{ijgmmp}), 
\begin{equation}\label{generalNUEq}
\psi''(x)+\frac{\Delta_{1}-\Delta_{2}x}{x(1-\Delta_{3}x)}\psi(x)+\left[\frac{-\zeta_{1}x^{2}+\zeta_{2}x-\zeta_{3}}{x^{2}\left(1-\Delta_{3}x\right)^2}\right]\psi(x)=0,
\end{equation}
can be expressed in terms of the generalized Jacobi polynomials $P_{n}^{(\alpha,\beta)}(x)$ as
\begin{equation}
\psi(x) = x^{\Delta_{12}}\left(1-\Delta_{3}x\right)^{-\Delta_{12}-\frac{\Delta_{13}}{\Delta_{3}}}P_{n}^{\left(\Delta_{10}-1, \frac{\Delta_{11}}{\Delta_{3}}-\Delta_{10}-1\right)}\left(1-2\Delta_{3}x\right).
\end{equation}
For the particular cases of Eq. \eqref{generalNUEq} where $\Delta_{3}=0$, the relevant wave functions are
\begin{equation}\label{WavefuncBeltrami}
\psi(x) = x^{\Delta_{12}} e^{(\Delta_{13}x)} \mathcal{L}^{\Delta_{10}-1}_{n} \left(\Delta_{11}x\right),
\end{equation}
where $\mathcal{L}^{\Delta}_{n}(x)$ is a generalized Laguerre polynomial.

The NU method also tells us that the energy eigenvalues of Eq. \eqref{generalNUEq} can be determined by the following equation:
\begin{equation}
\begin{split}
&n\Delta_{2}-(2n+1)\Delta_{5}+(2n+1)(\sqrt{\Delta_{9}}-\Delta{3}\sqrt{\Delta_{8}})+n(n-1)\Delta_{3}\\
&+\Delta_{7}+2\Delta_{3}\Delta_{8}+2\sqrt{\Delta_{8}\Delta_{9}} = 0,
\end{split}
\end{equation} 
where the symbols $\Delta_{1}$, $\Delta_{2}$, $\Delta_{3}$ and $\zeta_{1}$, $\zeta_{2}$, $\zeta_{3}$ are  the constants in Eq. \eqref{generalNUEq},  and
the remaining parameters $\Delta_{4}, \Delta_{5}, \dots \Delta_{13}$ can be calculated from $\Delta_{1}$, $\Delta_{2}$, $\Delta_{3}$, $\zeta_{1}$, $\zeta_{2}$, $\zeta_{3}$ by
\begin{equation}
\begin{split}
&\Delta_{4}=\frac{1}{2}\left(1-\Delta_{1}\right), \quad \Delta_{5}=\frac{1}{2}\left(\Delta_{2}-2 \Delta_{3}\right), \quad 
\Delta_{6}=\Delta_{5}^{2}+\zeta_{1}, \\
&\Delta_{7}=2 \Delta_{4} \Delta_{5}-\zeta_{2}, \quad \Delta_{8}=\Delta_{4}^{2}+\zeta_{3}, \quad 
\Delta_{9}=\Delta_{3}\Delta_{7}+\Delta_{3}^{2}\Delta_{8}+\Delta_{6},\\
&\Delta_{10}=\Delta_{1}+2\Delta_{4}+2\sqrt{\Delta_{8}}, \quad \Delta_{11}=\Delta_{2}-2\Delta_{5}+2\left(\sqrt{\Delta_{9}}+\Delta_{3}\sqrt{\Delta_{8}}\right), \\
&\Delta_{12}=\Delta_{4}+\sqrt{\Delta_{8}}, \quad \Delta_{13}=\Delta_{5}-\sqrt{\Delta_{9}}-\Delta_{3} \sqrt{\Delta_{8}}.
\end{split}
\end{equation}

From the above equations, we observe that Eq. \eqref{DiracEq7} is a special case of the NU differential equation with $\Delta_{3}=0$.  In order to solve Eq. \eqref{DiracEq7}, we apply the change of variable: 
\begin{equation}\label{q-u}
q=e^{-\frac{u}{r}}.
\end{equation} 
Thereby, the wave functions of the wave equation in
Eq. \eqref{DiracEq7}  would be the same Eq. \eqref{WavefuncBeltrami} with the following form
\begin{equation}\label{WavefuncBeltrami}
	\psi(q) = q^{\frac{1}{2}+\tau+\sqrt{(M^{2}-E^{BWH}_{nm})r^{2}}} e^{-\sqrt{\frac{m^{2}r^{2}}{b^{2}}}\,q} \mathcal{L}^{2\sqrt{(M^{2}-E^{BWH}_{nm})r^{2}}}_{n} \left(2\sqrt{\frac{m^{2}r^{2}}{b^{2}}}\,q\right),
\end{equation}
and the corresponding energy eigenvalues can be determined as
\begin{equation}\label{energy1}
	\begin{split}
		E^{BWH}_{nm}=\pm \sqrt{\frac{(1+2n)\left|m\right|r-mr\left[1+2n+2n^{2}-2M^{2}r^{2}\right]}{\sqrt{2m}\,r^{\frac{3}{2}}}}\,.
	\end{split}
\end{equation}
From Eq. \eqref{energy1}, we see that the energy  does not depend on the wormhole parameter $b$, but only on the wormhole parameter $r$, the quantum numbers $m$ and $n$, and the fermion's mass. Clearly, real values of energy are possible when $(1+2n)\left|m\right|r+2M^2mr^3\geq mr\left(1+2n+2n^2\right)$, which is more likely for large fermion masses and large values of the wormhole parameter $r$.

%%%%%%%%%%%%%%%%%%%%%%%%%%%%%%%%%%%%%%%%%%%%%%%%

%% Conclusion

%%%%%%%%%%%%%%%%%%%%%%%%%%%%%%%%%%%%%%%%%%%%%%%%

\section{Conclusion\label{sec7}}

We investigated the effects of an LSV and CPT-even non-minimal coupling on a neutral fermion in a $(1+2)$-dimensional fermion within the SME framework. The LSV is controlled by a constant background tensor field that is coupled to the fermions and the field-strength tensor which we reduced to one non-zero component such that our fermion became coupled to an external magnetic field in one (here, $x$) direction. Among our motivations are the SME-based predictions of minute experimental signatures in various systems, either on the Earth or in space; hence, in this paper, the effect on a neutral fermion in a wormhole-generated curved surface. The effects of this surface's curvature, to which a quantum fermion is confined, could lead to interesting consequences.  As discussed in Ref. \cite{RojjanasonEPJC1920} (the 2019 paper in EPJC), potential applications could be realized with curved graphene and nanotubes, who can play the role of a wormhole's throat bridging two graphene sheets, analogous to the wormhole's two surfaces. 

We introduced a modified Dirac Lagrangians for a massive neutral fermion subject with a covariant derivative that involved a non-minimal coupling leading to LSV and CPT-even electrodynamics.  We then set this theory on the curved space-time of a $(1+2)$-dimensional wormhole whose form we characterized by a function $R(u)$ of the meridian coordinate $u$ and whose constant Gaussian curvature can be written in terms of $R(u)$ and its second-derivative. As an intermediary step, we found a second-order differential equation \eqref{DiracEq4} in terms of $u$ and solved it for various expressions for $R(u)$ that described respectively hyperbolic, elliptic and Beltrami wormholes.

For the hyperbolic wormhole, we solved Eq. \eqref{DiracEq4} for two possible scenarios associated to two values of $\tau=-\mathrm{i}\lambda\left(k_{DB}\right)_{21}B_{0}$; that is, a parameter depending on the constant background tensor and the magnitude of the magnetic field. We found the wave-functions in terms of hypergeometric functions.   The structure of eigenfunctions and energies was analogous for the elliptic wormhole.  It was a little different for the Beltrami wormhole, for which we solved the differential equation by means of the Nikiforov-Uvarov method and found its wave-functions in terms of the generalized Laguerre polynomials. 

\section*{Data Availability Statement}

No data associated in the manuscript.

%%%%%%%%%%%%%%%%%%%%%%%%%%%%%%%%%%%%%%%%%%%%%%%%

%%Acknowledgement

%%%%%%%%%%%%%%%%%%%%%%%%%%%%%%%%%%%%%%%%%%%%%%%%

\section*{Acknowledgements}
%The authors thank the referee for.....
%%%%%%%%%%%%%%%%%%%%%%%%%%%%%%%%%%%%%%%%%%%%%%%%
The authors would like to thank Professor Fayçal Hammad, Bishop’s University, for many useful comments and suggestions.
%% REFERENCES

%%%%%%%%%%%%%%%%%%%%%%%%%%%%%%%%%%%%%%%%%%%%%%%%

\end{document}